\documentstyle[aps,multicol,psfig]{revtex}

\newcommand{\be}{\begin{equation}}
\newcommand{\ee}{\end{equation}}
\newcommand{\ba}{\begin{eqnarray}}
\newcommand{\ea}{\end{eqnarray}}

\begin{document}

\title{Robustness of Quadratic Solitons with Periodic Gain}

\author{Lluis Torner and Juan P. Torres}

\address{Laboratory of Photonics, Department of Signal Theory and 
Communications,\\
Universitat Politecnica de Catalunya,
Gran Capitan UPC-D3, Barcelona, ES~08034, Spain}

\author{Ole Bang}

\address{Department of Mathematical Modelling,
Technical University of Denmark, bldg.~321 DK-2800 Kgs.~Lyngby, Denmark}

\maketitle

\begin{abstract}
We address the robustness of quadratic solitons with periodic non-conservative
perturbations. We find the evolution equations for guiding-center solitons
under conditions for second-harmonic generation in the presence of periodic
multi-band loss and gain. Under proper conditions, a robust guiding-center
soliton formation is revealed.
\end{abstract}

\begin{multicols}{2}

\newlength{\figwidtha}
\newlength{\figwidthb}
\setlength{\figwidtha}{0.45\linewidth}
\setlength{\figwidthb}{0.95\linewidth}

\narrowtext

Multicolor optical soliton formation mediated by cascading of quadratic
nonlinearities has been demonstrated experimentally during the last few years
in a variety of geometries and frequency-mixing processes, in settings for
spatial, temporal and spatio-temporal trapping of light.$^1$ Soliton signals
exist in particular in the process of second-harmonic generation (SHG) that is
addressed here, where solitons form in waveguides and in bulk crystals by the
mutual trapping between the fundamental frequency and second-harmonic waves.
Multidimensional soliton families exist above a threshold light intensity for
all values of the phase-mismatch between the waves, and most of such
solitons have been shown to be dynamically stable under propagation with the
equations that model the ideal light evolution under conditions of
focused and pulsed SHG. Adiabatic soliton decay and amplification in the
presence of weak loss or gain, have been also studied.$^{2-4}$

In this Letter we address the robustness of quadratic solitons against 
strong, but periodic non-conservative perturbations. To start the program we
consider soliton formation in the presence of multi-frequency losses and large,
but rapidly-varying periodic gain. Our goal is to derive the corresponding
guiding-center evolution equations and to expose the robustness of the existing
solitons under proper conditions. We believe that the results reported bear a
generic fundamental interest to the robustness of quadratic soliton formation
in structures with periodic non-conservative perturbations. Moreover, they
might find direct applications in reduced models of multi-color laser systems
with intracavity frequency generation, including self-frequency doubling
schemes, operating in the solitonic regime.$^{5,6}$

Here the focus is on solitons formed in one-dimensional structures under
conditions for non-critical type I SHG, but the analysis can be extended to
different physical settings. The evolution of the slowly-varying envelopes of
the light waves in the presence of multi-frequency band loss and periodic gain
can be described by the reduced equations$^7$
\ba
 & &    i\frac{\partial a_1}{\partial\xi}
        -\frac{\alpha_1}{2}\frac{\partial^2 a_1}{\partial s^2}
        + a_1^*\, a_2\, \exp (-i\beta\xi)=i\Gamma_1(\xi)a_1,\\
 & &   i\frac{\partial a_2}{\partial\xi}
        -\frac{\alpha_2}{2}\frac{\partial^2 a_2}{\partial s^2}
         +a_1^2\, \exp (i\beta\xi)=i\Gamma_2(\xi)a_2,
\ea
where $a_1$ and $a_2$ are the normalized amplitudes of the fundamental
frequency (FF) and second-harmonic (SH) waves. In the case of spatial solitons,
$\alpha_1=-1$, and $\alpha_2=-k_1/k_2\simeq -0.5$, where $k_\nu$, with
$\nu=1,2$ are the linear wave numbers at both frequencies. In the case of
temporal solitons, $\alpha_\nu$ stand for the ratio between the
group-velocity-dispersions existing at both frequencies.  The 
transverse and longitudinal coordinates are normalized to the beam or pulse
width and to the diffraction or dispersion length at the fundamental frequency,
respectively. The parameter $\beta$ is the scaled phase mismatch, and 
$\Gamma_\nu(\xi)$ stand for periodic gain and loss. Let
$\Gamma_\nu(\xi)=\Gamma_\nu^{(0)} + R_\nu(\xi)$, where $\Gamma_\nu^0$ are the
average gain or loss at the FF and SH frequencies, and $R_\nu(\xi)$ are
periodic functions, with period $L\ll 1$, and zero mean.

To derive the evolution equations of the guiding-center solitons in the
presence of the periodic gain, we first use the approach originally
introduced by Mollenauer, Evangelides and Haus for the case of Kerr solitons
propagating in optical fibers.$^8$  Let $A_1=a_1$ and
$A_2=a_2\exp(-i\beta\xi)$.  The explicit periodic gain can be removed from
equations (1)-(2) by making the transformations
$A_\nu(\xi,s)=U_{\nu}(\xi,s)\exp[\int_0^\xi R_\nu(\xi')\mbox{d}\xi']$.
Substitution in (1)-(2) leads to the evolution equations
\ba
 & &    i\frac{\partial U_1}{\partial\xi}
        -\frac{\alpha_1}{2}\frac{\partial^2 U_1}{\partial s^2}
        -i\Gamma_1^{(0)}U_1
        + \epsilon_2(\xi) U_1^*\, U_2=0,\\
 & &   i\frac{\partial U_2}{\partial\xi}
        -\frac{\alpha_2}{2}\frac{\partial^2 U_2}{\partial s^2}
        -i\Gamma_2^{(0)}U_2-\beta\, U_2
         + \epsilon_3(\xi) U_1^2=0,
\ea
where the resulting longitudinally-varying nonlinear coefficients write
\ba
 & &    \epsilon_2(\xi)= \exp\left[\int_0^\xi R_2(\xi')\mbox{d}\xi'\right],\\
 & &    \epsilon_3(\xi)= \exp\left[
                \int_0^\xi \left(2R_1(\xi')-R_2(\xi')\right)\mbox{d}\xi'
                             \right].
\ea
To proceed further one now assumes that the wave evolution over a period of the
map is mostly dictated by the gain and loss, which are responsible
for fast amplitude oscillations of the fields $A_\nu$, in addition to the
residual effects induced by the nonlinearity. Therefore, $U_\nu$ vary
slowly over a period of the map, so that one can average equations (3)-(4) to
approximately get
\ba
 & &    i\frac{\partial U_1}{\partial\xi}
        -\frac{\alpha_1}{2}\frac{\partial^2 U_1}{\partial s^2}
        -i\Gamma_1^{(0)}U_1
        + \eta_2\, U_1^*\, U_2\simeq 0,\label{RN1}\\
 & &   i\frac{\partial U_2}{\partial\xi}
        -\frac{\alpha_2}{2}\frac{\partial^2 U_2}{\partial s^2}
        -i\Gamma_2^{(0)}U_2 -\beta\, U_2
         + \eta_3\, U_1^2\simeq 0,\label{RN2}
\ea
where
\be
        \eta_{2,3}=<\epsilon_{2,3}(\xi)>=
                \frac{1}{L}\,\int_0^L \epsilon_{2,3}(\xi)\mbox{d}\xi,
\ee
are the averaged nonlinear coefficients over a period of the map.
Therefore, the evolution equations for the slowly-varying averaged fields
$U_{\nu,0}(\xi,s)$, defined as 
\be
       U_{\nu,0}(\xi,s)=<\exp[\int_0^\xi R_\nu(\xi')\mbox{d}\xi']>U_\nu(\xi,s),
\ee
write
\ba
 & &    i\frac{\partial U_{1,0}}{\partial\xi}
        -\frac{\alpha_1}{2}\frac{\partial^2 U_{1,0}}{\partial s^2}
        - i\Gamma_1^{(0)}U_{1,0} 
        +  U_{1,0}^*\, U_{2,0}\simeq 0, 
        \label{GC1}\\
 & &   i\frac{\partial U_{2,0}}{\partial\xi}
        -\frac{\alpha_2}{2}\frac{\partial^2 U_{2,0}}{\partial s^2}
        -i\Gamma_2^{(0)}U_{2,0} -\beta\, U_{2,0}
         + \eta\, U_{1,0}^2\simeq 0,
        \label{GC2}
\ea
where $\eta=\eta_2\eta_3/\eta_1^2$, with $\eta_1=<\epsilon_1(\xi)>$, and
\be
     \epsilon_1(\xi) = \exp\left[ \int_0^\xi R_1(\xi')\mbox{d}\xi'\right].
\ee
One thus concludes that under the conditions where the approximations used to
derive (\ref{GC1})-(\ref{GC2}) hold, the guiding-center solitons are given by
those existing without gain and loss when $\Gamma_\nu^{(0)}=0$,$^9$ or
otherwise with compensated gain and loss,$^{10}$ but with properly
renormalized amplitudes. To expose the value and dependencies of the
renormalization coefficient $\eta$, consider the illustrative map
$R_\nu(\xi)=G_\nu\,H(\xi)$, where $H(\xi)$ is the Heaviside function
\be
\label{map}
        H(\xi)=\left\{\begin{array}{ccl}
        -1, &  & nL<\xi<(n+1/2)L,\\
         1, &  & (n+1/2)L<\xi<(n+1)L.
                      \end{array}\right.
\ee
Substitution in the above expressions and performing all the averages yields
\be
        \eta=\frac{F[G_2L/2]F[(2G_1-G_2)L/2]}{F^2[G_1L/2]},
        \label{eta}
\ee
where
\be
        F[\mu]=\frac{1-\exp(-\mu)}{\mu}.
\ee
Of particular interest are the weak maps with $\mu\ll 1$, corresponding to
$G_\nu L\ll 1$. In such cases, one might expand (\ref{eta}) in powers of the
corresponding parameters to get, at order $L^2$
\be
        \eta\simeq 1\,+\,\frac{(G_1-G_2)^2L^2}{48}.
        \label{etaapprox}
\ee
Note that when $G_1=G_2$, equation (\ref{eta}) gives $\eta=1$, regardless the
value of the map period $L$. However, this does not mean that the
guiding-center evolution is given by (\ref{GC1})-(\ref{GC2}) with $\eta=1$ at
all orders of $L$, because the derivation of (\ref{GC1})-(\ref{GC2}), hence the
governing equations themselves, are only intended to hold when $L\ll 1$.  To
elucidate the applicability limits of (\ref{GC1})-(\ref{GC2}), it is worth
deriving the guiding-center evolution equations using more mathematically
systematic approaches.$^{11,12}$ Next we outline the outcome of the asymptotic
expansion method developed by Kivshar and co-workers for similar problems but
for Kerr solitons.$^{12}$ Such approach has been recently employed to obtain
guiding-center evolution equations of light signals propagating in
quasi-phase-matched quadratic structures.$^{13}$

Assuming perfect periodicity of the functions $R_\nu(\xi)$ as above, one can
express the periodic gain and all fields as Fourier series of the form
\ba
        & & R_\nu(\xi)=\sum_{n\neq 0} d_{\nu,n}\exp(in\lambda\xi),\\
        & & A_\nu(\xi,s)=\sum_{n} U_{\nu,n}(\xi,s)\exp(in\lambda\xi),
\ea
where $\lambda=2\pi/L$ and the Foruier coefficients $U_{\nu,n}(\xi,s)$ vary
much slower than the corresponding carrier $\exp(in\lambda\xi)$. For the map
(\ref{map}) considered here one has
\begin{equation}
        d_{\nu,2n}=0,\qquad d_{\nu,(2n+1)}=\frac{2G_\nu}{i\pi(2n+1)}.
\end{equation}
Assuming the amplification maps to vary rapidly over a diffraction/dispersion
lenght, i.e.\ $L\ll 1$, so
that the spatial frequency $\lambda$ is large enough, one can expand the
harmonic amplitudes $U_{\nu,n\ne 0}$, as power series of $1/\lambda$.  Namely,
\be
  U_{\nu,n\ne 0}=\sum_{m>0} \frac{F_{\nu,n}^{(m)}}{\lambda^{m}}. 
\ee
Substitution in the governing equations and matching leading-order
contributions leads to
\be
        F_{\nu,n}^{(1)}=-i\,\frac{d_{\nu,n}}{n}\,U_{\nu,0}.
\ee
The guiding-center evolution is obtained by solving recursively for the
higher-order contributions $F_{\nu,n}^{(m>1)}$ and substituting the
corresponding expressions in the evolution equations for the average fields
$U_{\nu,0}$. At order $1/\lambda^2$, one exactly finds
(\ref{GC1})-(\ref{GC2}), with (\ref{etaapprox}).
        
Equations (\ref{GC1}), (\ref{GC2}) and (\ref{etaapprox}) are the central result
of this paper. First, they reveal that when $G_1=G_2$, at order $L^2$ the
guiding-center evolution equations are identical to those without the periodic
gain, regardless the value of the amplitudes $G_\nu$. This is a remarkable
result, that emphasizes the robustness of quadratic solitons with the type of
perturbations considered.  Second, the typical values of the gain coefficient
that is obtained in Erbium-doped lithium niobate or potassium titanyl phosphate
around the third telecommunication window centered at 1.55 $\mu$m fall in the
range 0-2 dB/cm.$^{5,14}$ In the case of spatial solitons, this yields values
of the normalized gain coefficient of the order of $G_\nu\sim 10^{-1}$.$^3$
With such values and letting $L\sim 10^{-1}$, one always obtains negligible
corrections of the order of $[\eta-1]\sim 10^{-5}$.

\begin{figure}
  \centerline{\hbox{\psfig{figure=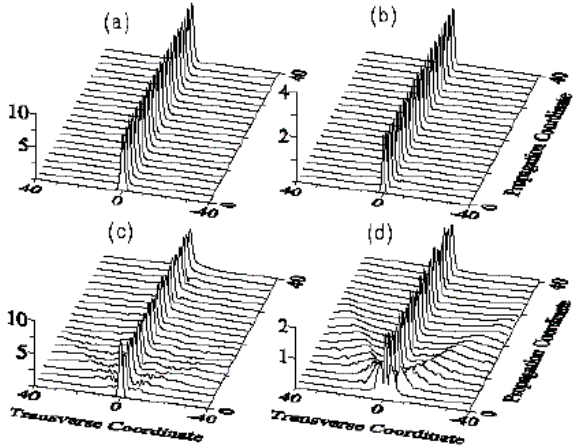,width=\figwidthb}}}
  \caption{Typical evolution of guiding-center quadratic solitons in 
  structures with rapidly-varying periodic gain. In (a)-(b): The input 
  is a renormalized stationary soliton solution. In (c)-(d): The input 
  contains only fundamental frequency light with a Gaussian shape. (a), 
  (c) : Evolution of the FF; (b), (d): Evolution of the SH. Conditions: 
  $\beta=3$, $\tilde{I}=30$, $G_1=20$, $G_2=0$, $L=0.2$.}
\end{figure}

To confirm that under proper conditions (\ref{GC1})-(\ref{GC2}) hold, we solved
(1)-(2) numerically for different input conditions and maps. Figure 1 shows
typical examples of the outcome, when $\Gamma_\nu^{(0)}=0$. The plots
correspond to the phase-mismatch $\beta=3$, but analogous results were obtained
for other values.  To emphasize that guiding-center solitons form with gigantic
values of the gain-loss amplitude, provided that the period of the map is small
enough so that the guiding-center approach is justified, we display results for
a map with $G_1=20$, $G_2=0$ and $L=0.2$.  Figures 1(a)-(b) show the
propagation of a guiding-center soliton excited by the input
$A_\nu(0)=\tilde{A}_\nu(0)/\rho_\nu$, where $\tilde{A}_\nu(0)$ are the
corresponding stationary solitons existing without gain and loss with energy
flow $\tilde{I}=\tilde{I}_1+\tilde{I}_2=\int (|\tilde{A}_1|^2+|\tilde{A}_2|^2)
\mbox{d}s=30$,$^9$ while $\rho_1=(\eta_2\eta_3)^{1/2}$, and $\rho_2=\eta_2$,
are the renormalization factors dictated by (\ref{RN1})-(\ref{RN2}). For the
map considered one has $\rho_1\sim 0.5$, so that in the renormalized input
the FF energy is enhanced by a factor of four. Figures 1(c)-(d) show the
excitation of a guiding-center soliton in the same map but with only FF input
light carrying the same energy flow as above in the Gaussian shape
$A_1(0)=(2/\pi)^{1/4}\,I_1^{1/2} \exp(-s^2)$. Here $I_1=\tilde{I}_1/\rho_1^2$.

\begin{figure}
  \centerline{\hbox{\psfig{figure=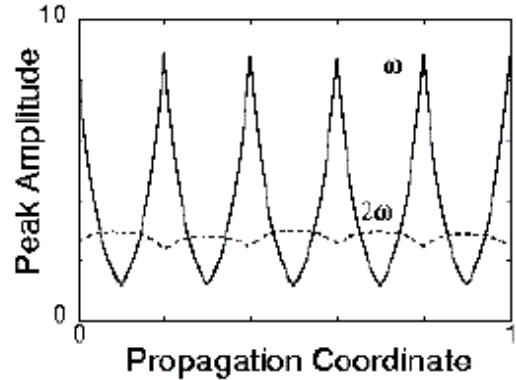,width=\figwidthb}}}
  \caption{Detail of the evolution of the peak amplitudes of the signals
  displayed in Figs.~1(a)-(b), over the first five periods of the map.}
\end{figure}

Figure 2 shows the detailed evolution of the field amplitudes in the case shown
in Figs.~1(a)-(b). Because of the large excursions of the amplitudes, the
actual evolution differs slightly from that predicted by equations
(\ref{GC1})-(\ref{GC2}). Such departures are responsible, e.g., of the small
resonance peaks appearing in the Fourier spectra of the FF and SH evolving
signals, as shown in Figure 3.

\begin{figure}
  \centerline{\hbox{\psfig{figure=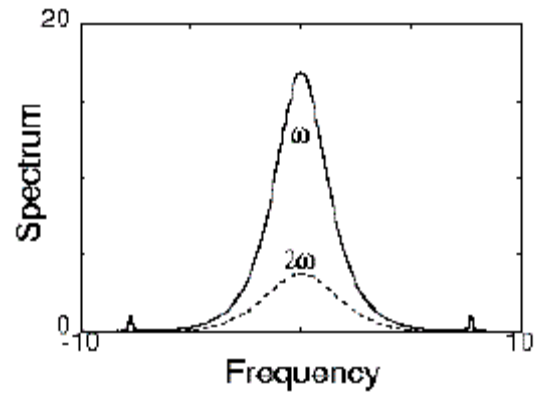,width=\figwidthb}}}
  \caption{Fourier spectra of the FF and SH signals displayed in
  Figs.~1(a)-(b), at $\xi=40$.}
\end{figure}

When $\Gamma_\nu^{(0)}\ne 0$, but the net loss at one frequency is compensated
for by the presence of a net gain at the other frequency band, the system
(\ref{GC1})-(\ref{GC2}) allows stationary soliton solutions,$^{10}$ which
are chirped. However, an interesting result revealed by the guiding-center
evolution equations (\ref{GC1})-(\ref{GC2}) is that, in the absence of a net
gain or loss (i.e., $\Gamma_\nu^{(0)}=0$), on average the guiding-center
quadratic solitons are chirpless. Accordingly, in contrast to what is found
with solitons in other periodic systems, e.g., with dispersion-managed
solitons propagating in optical fibers,$^{15}$ one concludes that with
chirpless input conditions the guiding-center quadratic solitons with periodic
gain are best excited when the first domain has the whole nominal length. The
outcome of our numerical simulations confirms that such is indeed the case, as
shown in Figure 4. The plot shows the evolution over a period of the map of the
quantities 
\be
        C_\nu(\xi)=\frac{i}{4}\,
              \frac{\int s (A_\nu A_{\nu,s}^* - A_\nu^* A_{\nu,s})\,\mbox{d}s} 
              {\int s^2 |A_\nu|^2\, \mbox{d}s},
\ee
where $A_{\nu,s}$ stands for $\partial A_\nu/\partial s$. The evolution
displayed corresponds to a guiding-center soliton similar to that shown in
Figs.~1(a)-(b), but for $G_1=G_2=10$.

\begin{figure}
  \centerline{\hbox{\psfig{figure=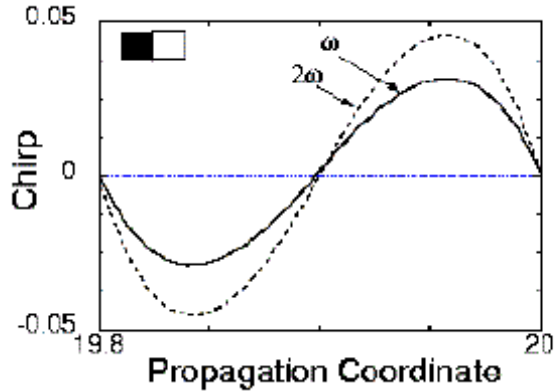,width=\figwidthb}}}
  \caption{Typical evolution of the integral chirp of the guiding-center
  solitons over a period of the map. The plot shows the evolution from
  $\xi=19.8$ to $\xi=20$. Conditions: $\beta=3$, $\tilde{I}=30$, $G_1=10$,
  $G_2=10$, $L=0.2$. The input is the corresponding renormalized stationary
  soliton solution.}
\end{figure}

\begin{figure}
  \centerline{\hbox{\psfig{figure=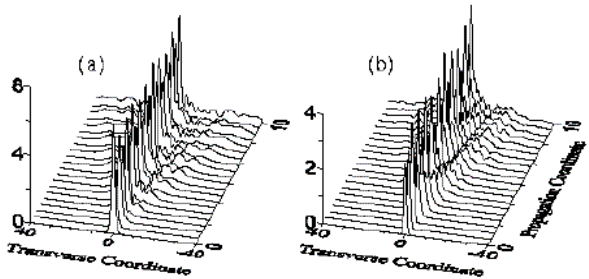,width=\figwidthb}}}
  \caption{Unstable evolution when the guiding-center approach fails.  
  The input is a renormalized stationary soliton solution. Same conditions 
  as in Figs.~1(a)-(b), except that here $G_1=2$ and $L=1$. (a): Evolution 
  of the FF; (b): Evolution of the SH.}
\end{figure}

Analogous results than those shown above were obtained for a variety of values
of the phase-mismatch $\beta$, the input energy flow $I$, the signal shape, and
the map amplitude. Naturally, this is so provided that the input conditions,
the gain amplitude, and the period $L$ make a guiding-center approach
justified.$^{11-12}$ Otherwise, e.g., when $L$, $G_\nu$ or the input $I$ are
too large, the guiding-center evolution fails, yielding a totally new scenario.
In particular, under such conditions resonance phenomena that make the wave
propagation unstable can occur. Figure 5 shows a typical example. The plot
shows the unstable evolution of the renormalized solution with $\tilde{I}=30$
in a map with $G_1=2$, with $L=1$.

In conclusion, the evolution equations for guiding-center quadratic solitons
propagating in structures with multi-frequency losses and rapidly-varying
periodic gain have been presented. Under proper conditions, robust multicolor
quadratic soliton formation has been revealed. Results might find applications
to reduced models of multicolor laser systems with intracavity frequency
generation, including self-frequency doubling schemes, operating in the
solitonic regime. Extension of the analysis to general maps that include the
details of the laser structures, including the pump light-matter
interaction,$^{5,14}$ is worth investigating.

This work was supported by the Generalitat de Catalunya. Ole Bang acknowledges
support from the Danish Technical Research Council through Talent Grant
No.~9800400.

\end{multicols}

\end{document}